\documentclass[aps,prd,twocolumn,groupedaddress,showpacs]{revtex4}

\newcommand{\pslash}{D\kern-0.15em\raise0.17ex\llap{/}\kern0.15em\relax}
\begin{document}

\title{Black holes and Rindler superspace: classical singularity and quantum unitarity}
\author{Chung-Hsien Chou$^{1}$,
Chopin
Soo$^{2}$
and
Hoi-Lai Yu$^{1}$
}
\affiliation{$^1$ Institute of Physics, Academia Sinica,
Nankang, Taipei 11529, Taiwan\\
$^2$Department of Physics, National Cheng Kung University, Tainan,
Taiwan}
\begin{abstract}

Canonical quantization of spherically symmetric initial data
appropriate to classical interior black hole solutions in four
dimensions is solved exactly without gauge-fixing the remaining
kinematic Gauss Law constraint. The resultant mini-superspace
manifold is two-dimensional, of signature $(+, -)$, non-singular,
and can be identified with the first Rindler wedge. The associated
Wheeler-DeWitt equation with evolution in intrinsic superspace
time is a free massive Klein-Gordon equation; and the
Hamilton-Jacobi semiclassical limit of plane wave solutions can be
matched to the interiors of Schwarzschild black holes. Classical
black hole horizons and singularities correspond to the boundaries
of the Rindler wedge. Exact wavefunctions of the Dirac equation in
superspace are also considered. Precise correspondence between
Schwarzschild black holes and free particle mechanics in
superspace is noted. Despite the presence of classical
singularities, hermiticity of the Dirac hamiltonian operator, and
thus unitarity of the quantum theory, is equivalent to an
appropriate boundary condition which must be satisfied by the
quantum states. This boundary condition holds for quite generic
quantum wavepackets of energy eigenstates, but fails for the usual
Rindler fermion modes which are eigenstates with zero uncertainty
in energy.

\end{abstract}
\pacs{04.60.Ds, 04.60.-m}
\maketitle
 \section{Introduction and overview}

 Quantization of the
spherically symmetric sector of four-dimensional pure General
Relativity(GR) has been investigated in a number of
articles\cite{Na,HR,AB}. Despite the simplicity of this toy model,
it is an extremely interesting testing (toy)ground for many
intriguing and challenging quantum gravity issues, both on the
technical as well as conceptual fronts. Birkhoff's
theorem\cite{Birkhoff} ensures that the classical solutions of
this sector are none other than Schwarzschild black holes, and as
such they come with classical horizons and singularities. How do
these manifest themselves and what roles do they play in the
quantum context? Do they affect the unitarity of the quantum
theory, and if so in what explicit manner? And for
reparametrization invariant theory, one ought to ask whether the
unitarity or non-unitary of quantum evolution is with respect to
intrinsic time in superspace\cite{DeWitt}, or some other
appropriate choice of ``time". Indeed it is not unreasonable to
demand any respectable quantum theory of gravity to provide
guidance on these issues.

Recent investigations motivated by loop quantum gravity have
yielded as insights and possible resolutions an upper bound on the
curvature, and reformulation of the Wheeler-DeWitt constraint as a
difference time evolution equation wherein classical black hole
singularities are not obstacles\cite{AB,HR}. Here we adopt a more
conservative approach based upon continuum physics and exact
canonical quantization. The Wheeler-DeWitt
equation\cite{DeWitt,Wheeler} for spherically symmetric
mini-superspace appropriate to the discussion of interior black
hole solutions\cite{AB} is solved exactly. Among the neat features
which emerged are: 1)The resultant arena for quantum
geometrodynamics is two-dimensional, of signature $(+, -)$,
non-singular, and can be identified with the first Rindler wedge.
2)Classical black hole horizons and singularities correspond,
remarkably, to the boundaries of the Rindler wedge. 3)The
classical super-Hamiltonian constraint is equivalent to a massive
free-particle dispersion relation in flat superspace; and the
Wheeler-DeWitt equation is a Klein-Gordon equation with evolution
in intrinsic superspace time. 4)The Hamilton-Jacobi semiclassical
limit consists of plane wave solutions which can be matched to the
interiors of Schwarzschild black holes. 5)Positive-definite
``probability" current for wavefunctions and other considerations
motivate the investigation of ``fermionic" solutions of the
associated Dirac equation. 6)With respect to the natural inner
product, hermiticity of the Dirac hamiltonian, and thus unitarity
of the quantum theory, is equivalent to a condition at the
boundary of space-like hypersurfaces of the Rindler wedge which
must be satisfied by the solutions. 7)The boundary condition is
demonstrated to be satisfied by rather generic wavepackets, but
not for the usual energy-eigenstate fermionic Rindler modes. This
situation has an analogy in free non-relativistic Schrodinger
quantum mechanics wherein hermiticity of the momentum operator
requires the physical Hilbert space to be made up of states which
are suitable wavepackets which vanish at spatial infinity, and
rule out plane wave states with infinitely sharp momentum. From
this perspective, the quantum evolution here is thus physically
unitary; and the boundary condition guaranteeing unitarity is much
milder than, say, ``Quantum Censorship" requirement of vanishing
wavefunction at the boundary.

\section{Spherically symmetric initial data and canonical quantization}

We start by following Ref.\cite{AB}, wherein modulo gauge
invariance, the Lie-algebra-valued connection 1-form for
spherically symmetric configurations is
\begin{eqnarray}
A &=& c T_3 d\alpha + (a T_1 + b T_2)d \theta +(- b T_1 + a T_2)
\sin{\theta} d \phi \nonumber \\ && + T_3 \cos{\theta} d \phi,
\label{e:111}
\end{eqnarray}
where $a, b, c$ are constants on space-like Cauchy surfaces, and
$T_a$ are the generators of $SO(3)$. The most general spherically
symmetric gauge configurations\cite{Witten} allow for $\alpha$
dependence as well, resulting in additional contributions to the
constraints for initial data relevant to exterior black hole
solutions. The description of exterior regions wherein
$\alpha$-dependence of $a, b$ and $c$ gives rise to further
contributions to the constraints has also been discussed elsewhere
recently\cite{Pullin}. As in Ref.\cite{AB}, the formalism here
works perfectly for the purposes of discussing the classical
singularity and properties inside the horizon.
The corresponding dreibein are
$e_1 = \omega_a d \theta - \omega_b \sin{\theta} d \phi, e_2 =
\omega_b d \theta + \omega_a \sin{\theta} d \phi, e_3 = \omega_c
d\alpha$; and they are related to the densitized triad by ${\tilde
E} = p_c T_3 \sin{\theta} \frac{\partial}{\partial \alpha} + (p_a
T_1 + p_b T_2) \sin{\theta} \frac{\partial}{\partial \theta} +(
p_a T_2 - p_b T_1 )\frac{\partial}{\partial \phi}$, with
$\omega_a = \frac{\sqrt{\mid p_c \mid} p_a}{\sqrt{p_a^2 +
p_b^2}},\quad \omega_b = \frac{\sqrt{\mid p_c \mid} p_b}{\sqrt{p_a^2
+ p_b^2}},\quad \omega_c = \frac{sgn {(p_c)}\sqrt{p_a^2 +
p_b^2}}{\sqrt{\mid p_c \mid} }$.


In this formulation, the Ashtekar-Barbero super-Hamiltonian, with
Immirzi parameter $\gamma$, is proportional to $\int d^3 x N
e^{-1} \left[ \epsilon^{abc} F_{ija}{\tilde E}^i_b{\tilde E}^j_c -
2 (1+\gamma^2) K^a_{[i} K^b_{j]}{\tilde E}^i_a{\tilde E}^j_b
\right]$, wherein $e := \sqrt{| \det{\tilde E}|} \, {\rm sgn}(\det
{\tilde E})$. Using the extrinsic curvature $K
=\frac{1}{\gamma}(A-\Gamma)$, where $\Gamma$ is the torsionless
connection compatible with the dreibein, the super-Hamiltonian
constraint simplifies to
\begin{eqnarray}
{\cal H} \propto 2 p_cc(p_aa + p_bb) + (p_a^2 + p_b^2) (a^2 + b^2
+ \gamma^2) \approx 0;
\end{eqnarray}
and the symplectic form for the spherically symmetric sector
is\cite{AB} $\Omega = \frac{1}{2 \gamma G}( 2 \delta a \wedge
\delta p_a + 2 \delta b \wedge \delta p_b + \delta c \wedge \delta
p_c)$.

To tackle gauge invariance, we define $p_a =: r \cos\Theta$, $p_b
=: r \sin\Theta$, $p_c :=: z$. Invoking the canonical quantization
rules, $\frac{2a}{2\gamma G} = \frac{\hbar}{i}{{\partial} \over
{\partial{p_a}}}=\frac{\hbar}{i}(
\cos\Theta\frac{\partial}{\partial r}
-\frac{\sin\Theta}{r}\frac{\partial}{\partial\Theta});
\frac{2b}{2\gamma G} = \frac{\hbar}{i}{{\partial} \over
{\partial{p_b}}}=\frac{\hbar}{i}(
\sin\Theta\frac{\partial}{\partial r}
+\frac{\cos\Theta}{r}\frac{\partial}{\partial\Theta});
\frac{c}{2\gamma G} =: p_z = \frac{\hbar}{i}
\frac{\partial}{\partial z}$; the quantum constraint acting on
wavefunctions $\Phi$ is thus
\begin{eqnarray}
\left [z{\hat p_z}r{\hat p_r} -
\frac{\hbar^2r^2}{4}(\frac{1}{r}\frac{\partial}{\partial
r}r\frac{\partial}{\partial r} +
\frac{1}{r^2}\frac{\partial^2}{{\partial\Theta}^2}) + \frac{r^2}{4
G^2} \right ] \Phi =0.
\end{eqnarray}
The Gauss law constraint $p_ba -  p_ab = 0$ is precisely
equivalent to invariance under $\Theta$-rotations about the
z-axis, which is solved exactly, without gauge-fixing, by the
independence of $\Phi$ with respect to $\Theta$. For the case of
$\alpha$-independent variables discussed here, the diffeomorphism
supermomentum constraint gives rise to no further restrictions.
Although $r$ and $z$ are the resultant superspace variables, it
can be readily verified the resultant Wheeler-DeWitt equation,
\begin{eqnarray}
\left[ - z \frac{\partial}{\partial z} r  \frac{\partial}{\partial
r} - \frac{1}{4}r  \frac{\partial}{\partial r}r
\frac{\partial}{\partial r} + \frac{r^2}{4 \hbar^2 G^2}\right ]
\Phi(r,z) =0, \label{e:177}
\end{eqnarray}
is not separable in $r$ and $z$.

To discuss the complete set of solutions, another set of variables
$X_- := \sqrt{z}; \quad X_+ := \frac{r^2}{4\hbar^2G^2\sqrt{z}}$ is
crucial. The Wheeler-DeWitt constraint in superspace now reduces
to
\begin{equation} X_+X_- \left [ -\frac{\partial}{\partial
X_+}\frac{\partial}{\partial X_-} + 1 \right ]\Phi(X_+,X_-) = 0,
\end{equation}
which is equivalently a simple Klein-Gordon wave equation in light
cone coordinates $X_\pm = \frac{1}{2}(X \pm T)$ with unit ``mass".
Plane waves $\Phi(X_+,X_-) = \exp(ik_+ X_+ + ik_- X_-)$ (with mass
shell condition $k_+k_- + 1 = 0$) and their superpositions are all
solutions, and the semiclassical limit of the theory will be
discussed in the following section.

\section{Semiclassical limit, Hamilton-Jacobi equations, and Schwarzschild black holes}

The semiclassical limit can be addressed by considering usual
semiclassical states $Ae^{i\frac{S}{\hbar}}$; with $A$ being a
slow varying function, and $S$ satisfying the Hamilton-Jacobi
equation, $\frac{\partial S}{\partial X_+}\frac{\partial
S}{\partial X_-} + 1 = 0$. Separation of variables with $S(X_+,
X_-) = S_+(X_+) + S_-(X_-)$ leads to $\frac{dS_\pm(X_\pm)}{dX_\pm}
= p_\pm$ being constants related by $p_+ = -1/p_-$; thus yielding
(up to a constant) the Hamilton function $S(X_+, X_-) = p_+X_+ +
p_-X_- $. These semiclassical states may thus be identified with
plane waves $A'e^{i(k_+X_+ +k_-X_-)}$ wherein $p_\pm = \hbar
k_\pm$.

We can furthermore map the equations of motion(EOM) for plane wave
solutions to classical black holes by studying the correspondence
of the classical initial data to the Hamilton-Jacobi theory. To
wit, we note that interior to the horizon, $(0 < R < 2GM)$, the
Schwarzschild metric with constant-$R$ Cauchy surfaces can be
described by the vierbein $e^A$ =
$(\frac{dR}{\sqrt{(\frac{2GM}{R}-1)}}, \frac{R}{\sqrt{2}}(d\theta
-\sin\theta d\phi), \frac{R}{\sqrt{2}}(d\theta +\sin\theta d\phi),
\sqrt{(\frac{2GM}{R}-1)}\, dt).$ Computing the Ashtekar potential
from $A_a = \gamma\omega^0\,_a
-\frac{1}{2}\epsilon_{abc}\omega^{bc}$, where $\omega_{AB}$ is the
torsionless spin connection, and comparing directly with
Eqs.(\ref{e:111}) and the corresponding dreibein yield,
\begin{eqnarray}
\omega_a = \omega_b =\frac{R}{\sqrt{2}},\quad \omega_c d\alpha =
\frac{r}{\sqrt{z}}d\alpha = \sqrt{(\frac{2GM}{R}-1)} dt;
\end{eqnarray}
\begin{eqnarray}
a = b = \frac{\gamma}{\sqrt{2}}{\sqrt{(\frac{2GM}{R}-1)}}, \quad c\,
d\alpha =-\gamma\frac{GM}{R^2}dt.
\end{eqnarray}
It follows that $X^2_- \equiv z = p_c = \omega_a^2 + \omega_b^2 =
R^2$. To infer the EOM from the Hamilton-Jacobi theory, we note that
on plane wave states, $\Phi \propto e^{(ik_+X_+ + ik_-X_-)}$,
\begin{equation}\hat{p_r}\Phi =
(\frac{2X_+\hbar k_+}{r})\Phi;\,\,  \hat{p_z}\Phi = (\frac{k_-X_-
- k_+X_+}{2X^2_-})\hbar\Phi. \label{e:333}
\end{equation}
Direct substitutions into the semiclassical Hamilton-Jacobi limit
of Eq.(\ref{e:177}) yield, as expected, the ``straight-line
trajectory" EOM,
\begin{equation}
X_+ = \frac{1}{k^2_+}({2GM} - X_-),
\end{equation}
which is independent of $\gamma$. Furthermore,
$\frac{r}{\sqrt{z}}\,d\alpha = \sqrt{(\frac{2GM}{R}-1)}dt$ gives
$(\frac{d\alpha}{dt}) = \frac{k_+}{2G\hbar}$ on applying the EOM.

\section{Rindler superspace}

 It should be noted that $X_\pm \geq
0$; thus the associated superspace is not the whole of Minkowski
spacetime; rather, it is precisely the first Rindler wedge, which
can be parametrized by $(\xi \geq 0, -\infty < \tau < \infty)$,
with $X = \xi\cosh\tau$ and $T =\xi\sinh\tau$, and endowed with
supermetric $ds^2 = dT^2 - dX^2 = \xi^2d\tau^2 - d\xi^2$.

Our previous EOM implies that {\it classical} black hole horizons
and singularities occur at $(X_- = R = 2GM,\, X_+ =
\frac{1}{k^2_+}[{2GM} - X_-] = 0)$ and $(X_- = 0, X_+ =
\frac{1}{k^2_+}[{2GM} - X_-] = \frac{2GM}{k^2_+})$ respectively.
Thus as we span over the range of black hole masses $M$, we
observe that the lower boundary $(X_- = 2GM, X_+ = 0)$ and upper
boundary $(X_- = 0,\frac{2GM}{k^2_+})$ of the Rindler wedge
correspond precisely to the classical black hole horizons and
singularities.

On this Rindler wedge the Klein-Gordon equation is
\begin{equation}
\left [\frac{1}{\xi^2}\frac{\partial^2}{{\partial
\tau}^2}-\frac{\partial^2}{{\partial
\xi}^2}-\frac{1}{\xi}\frac{\partial}{{\partial \xi}} + 1 \right
]\Phi(\xi,\tau) =0. \label{e:KG}
\end{equation}
The orthonormal modes can be expressed explicitly as
$\Phi_\omega(\xi,\tau) =
\sqrt{\frac{1-e^{-{2\pi\omega}}}{2}}B_\omega(\xi,\tau)$, wherein
the ``Minkowski Bessel modes"\cite{Gerlach} are given by $B_\omega
=
\frac{1}{\pi}e^{\frac{\pi\omega}{2}}K_{i\omega}(\xi)e^{-i\omega\tau}$.
These Minkowski Bessel modes can in fact be understood as the
``rapidity Fourier transform" of plane wave solutions $e^{i(kX
-\omega T)}$\cite{Gerlach} i.e. $B_\omega(\xi,\tau) =
\frac{1}{2\pi}\int^{\infty}_{-\infty}e^{i(\xi\sinh(\eta-\tau))}
e^{-i\omega\eta}d\eta$; where $k_\pm = k \mp \omega$, and in terms
of $\eta$ (the rapidity variable), $k = \sinh\eta, \omega =
\cosh\eta$, thus satisfying the mass shell condition $\omega^2 =
k^2 + 1$ for our particular case of unit mass. As discussed, the
plane wave solutions which correspond to Schwarzschild black holes
are given by the inverse rapidity Fourier transforms of
$B_\omega$. A conserved current with respect to the Klein-Gordon
inner product can be constructed\cite{DeWitt}; however, it is
well-known in Klein-Gordon theory positive-definite current
density for generic superpositions cannot be guaranteed. 

While ``bosonic" scalar solutions of Eq.(\ref{e:KG}) are the first
to come to mind, the possibility of ``fermionic" solutions of the
Wheeler-DeWitt constraint should not be ignored. To wit, we
investigate the Dirac equation which is {\it first-order in
intrinsic superspace time} on the Rindler wedge,
\begin{equation}
(i\pslash - m)\Psi = [\frac{\gamma^0}{\xi}i\partial_\tau +
\frac{i\gamma^1}{2\xi} +{i}\gamma^1\partial_\xi - m]\Psi = 0,
\label{e:DE}
\end{equation}
with $\Omega$ being the spin connection compatible with Rindler
space vielbein ${\textsc{e}}^{0,1}\,\,_\mu dx^\mu = (\xi d\tau,
d\xi)$. Moreover, the Lorentz-invariant current density ${\tilde
J}^\mu \equiv det({\textsc{e}}){\overline\Psi}\gamma^\mu\Psi$
obeys $\partial_\mu{\tilde J}^\mu = 0 $ and ${\tilde J}^0
=det(\textsc{e})\textsc{e}^0\,_A {\overline\Psi}\gamma^A\Psi =
\Psi^\dagger\Psi$ is also positive-definite. Orthonormal modes
with respect to the inner product $\langle\Psi'|\Psi\rangle
=\int^\infty_0 \Psi'^\dagger\Psi d\xi$ can be expressed
as\cite{Oriti, CD} $\Psi_\nu(\tau,\xi) =N_\nu
e^{-i\nu\tau}K_{i\nu -({\gamma^0\gamma^1}/{2})}(\xi)\chi$. 
It is possible to choose $\chi = \chi^+ + \chi^-$, with constant
orthonormal $\pm$-eigenspinors, $\chi^\pm$, of $\gamma^0\gamma^1$
with the resultant normalization constant $N_\nu =
\frac{\sqrt{\cosh(\pi\nu)}}{\pi}$.

\section{Quantum unitarity despite the presence of apparent classical singularities}

 With respect to the inner product discussed
above, the Dirac hamiltonian operator on the Rindler wedge, $H_D =
-{i}\gamma^0\gamma^1\xi\partial_\xi -\frac{i\gamma^0\gamma^1}{2} +
m\gamma^0$, is hermitian  i.e. $\langle H_D\Psi'|\Psi\rangle =
\langle\Psi'|H_D\Psi\rangle$ iff for all $-\infty < \tau <
\infty$, $\lim_{\xi \rightarrow
0}\xi\Psi'^\dagger(\tau,\xi)\gamma^0\gamma^1\Psi(\tau,\xi) =
0$\footnote{The boundary of space-like hypersurfaces for $-\infty
< \tau < \infty$ are at the origin with $\xi=0$, and at $\xi
=\infty$ where the fall-off behavior of $K_{i\nu \pm \frac{1}{2}}$
results in no further restrictions. The hypersurfaces $\xi =0,
\tau =\pm\infty$ are light-like in superspace rather than
space-like.}. In an investigation of the Unruh effect,
Oriti\cite{Oriti} conjectured that the fermionic solutions should
therefore obey $\lim_{\xi \rightarrow 0}
\xi^{\frac{1}{2}}\Psi(\xi) = 0$. In our present context of quantum
gravity, it is intriguing this last equation is a boundary
condition which guarantees hermiticity of the hamiltonian and thus
{\it unitarity of the quantum evolution with respect to intrinsic
superspace time} despite the threat of apparent classical black
hole singularities.

The asymptotic behavior of $K_\alpha(\xi)$ is quite simple:
\begin{equation} \lim_{\xi \rightarrow 0}\xi^{\frac{1}{2}}K_{i\nu \pm
\case{1}{2}}(\xi) =\sqrt{2}\,\Gamma(\case{1}{2} \pm
i\nu)({\case{\xi}{2}})^{\mp{i\nu}},
\end{equation}
where $\pm\vartheta(\nu)= Arg[\Gamma(\case{1}{2} \pm i\nu)]$.
Since $\{\Psi_\nu(\tau,\xi)\}$ forms an orthonormal basis, the
boundary condition for unitarity implies the restriction (on
$f(\nu)$) on a generic state $\Psi(\tau,\xi)$ which is
\begin{eqnarray} 0 &=&\lim_{\xi \rightarrow
0}\xi^{\case{1}{2}}\Psi(\tau,\xi) = \lim_{\xi \rightarrow
0}\sqrt{\xi}\int^\infty_{-\infty} d\nu f(\nu)\Psi_\nu(\tau,\xi)\cr
&=& \sqrt{\case{2}{\pi}}\lim_{\xi \rightarrow
0}\int^\infty_{-\infty} d\nu\, f(\nu)e^{-i\nu\tau}\times\cr &&
 [e^{i\vartheta(\nu)}(\case{\xi}{2})^{-i\nu}\chi^+
 + e^{-i\vartheta(\nu)}(\case{\xi}{2})^{i\nu}\chi^-].
\end{eqnarray} This results in
two conditions:
\begin{equation}\lim_{X_\pm \rightarrow 0}\int^\infty_{-\infty}
d\nu\, f(\nu)
 e^{\pm{i}\vartheta(\nu)}e^{\mp{i}\nu\ln X_\pm} = 0;\end{equation}
wherein $X_\pm = \frac{\xi}{2} e^{\pm\tau}$ has been used, and for
$ -\infty < \tau < \infty$, the limit $\xi \rightarrow 0$ implies
both $X_\pm \rightarrow 0$. Thus the boundary condition is
equivalent to requiring the Fourier transform of $f(\nu)e^{\pm
i\vartheta(\nu)}$ to vanish at $\pm\infty$. Since
$e^{i\vartheta(\nu)}$ is oscillatory (rather than sharply peaked),
the condition can be satisfied by a rather generic wavepacket with
$f(\nu)$ whose Fourier transform vanishes at $\pm\infty$
(explicitly, for instance, by $f(\nu)$ being Gaussian). From this
perspective, the quantum evolution here is thus physically
unitary. However, it should be pointed out that infinitely sharp
energy eigenstates i.e. $f(\nu_0) \propto \delta(\nu-\nu_0)$ which
correspond to the usual Rindler modes $\Psi_{\nu_0}$ {\it fail} to
satisfy the boundary condition. An analogous situation happens in
free non-relativistic quantum mechanics wherein hermiticity of the
momentum operator requires a physical Hilbert space of suitable
wavepackets which vanish at spatial infinity, and rule out plane
wave eigenstates with infinitely sharp momentum.

It can also be verified that a semiclassical black hole state is a
wavepacket (in $\nu$) which satisfies the boundary condition
explicitly. We note that the energy eigenstates
$\Psi_\nu(\tau,\xi)$ satisfy the associated equation,
\begin{eqnarray}0=(i\pslash + m)(i\pslash - m)\Psi_\nu\qquad\qquad\qquad&&\cr
=[-\frac{1}{\xi^2}e^{-(\frac{\gamma^0\gamma^1\tau}{2})}\partial^2_\tau
e^{(\frac{\gamma^0\gamma^1\tau}{2})}+\partial^2_\xi+\frac{1}{\xi}\partial_\xi-m^2]\Psi_\nu.&&
\end{eqnarray}
This means that if $\Psi$ is a solution of the Dirac equation, the
fermionic Lorentz-boosted solution $\Phi^F(\tau,\xi) =
e^{\frac{\gamma^0\gamma^1\tau}{2}}\Psi(\tau,\xi)$ will also solve
the Klein-Gordon equation (\ref{e:KG}).
Spinorial Minkowski plane wave modes can be written (in terms of
the rapidity $\eta$),
as $P^F_\eta(X,T) =
\frac{1}{2\sqrt{2\pi}}(e^{\frac{\eta}{2}}\chi^- +
ie^{-\frac{\eta}{2}}\chi^+)e^{i(kX-\omega T)}$, with
orthonormalization $\int^\infty_{-\infty} (P^F_{\eta'})^\dagger
P^F_\eta dX = \delta(\eta-\eta')$.
The identity $K_{i\nu\mp\frac{1}{2}}e^{-(i\nu\mp\frac{1}{2})\tau}
= \frac{1}{2}e^{i\frac{\pi}{2}(i\nu\mp\frac{1}{2})}
\int^\infty_{-\infty}e^{-(i\nu\mp\frac{1}{2})\eta}
e^{i(kX-\omega{T})}d\eta$ implies that the Rindler energy
eigenstates, $\Psi_\nu(\tau,\xi) = N_\nu e^{-i\nu\tau}[K_{i\nu
-\frac{1}{2}}(\xi)\chi^+ + K_{i\nu + \frac{1}{2}}(\xi)\chi^-]$,
and Minkowski plane wave states are thus related by
$e^{\frac{\gamma^0\gamma^1\tau}{2}}\Psi_\nu(\tau,\xi) =
\frac{N_\nu}{2}e^{-\frac{\pi\nu}{2}}\int^\infty_{-\infty}
 P^F_\eta(T,X)e^{-i\nu\eta}d\eta$.
Thus the semiclassical fermionic plane wave black hole
solution which solves the Dirac equation (\ref{e:DE}) is just the
inverse Fourier transform
\begin{equation}
e^{-\frac{\gamma^0\gamma^1\tau}{2}}P^F_\eta(T,X)
=2\pi\int^\infty_{-\infty}(\frac{2}{N_\nu}e^{\frac{\pi\nu}{2}})\Psi_\nu(\tau,\xi)e^{i\nu\eta}
d\nu
\end{equation}
which is a {\it wavepacket} of Rindler modes $\Psi_\nu$ with
$f_\eta(\nu)= \frac{4\pi}{N_\nu}e^{(i\eta +\frac{\pi}{2})\nu}$.

\section{Summary and Further remarks}

 The limitations of classical GR are
saliently exposed by the occurrence of singularities in the
theory. While singular potentials do not necessarily pose problems
to quantum theory (in quantum mechanics situations with singular
potentials are tractable and even exactly solvable), classical
curvature singularities need not be manifested in the quantum
context as singular potentials. The quantum theory of spherically
symmetric 4-dimensional GR is a simple model whose classical
sector is made up of only black hole solutions to Einstein's
theory. In this sense it is precisely a ``quantum theory of
Schwarzschild black holes", just as non-relativistic free particle
Schrodinger quantum mechanics is a quantum theory of free
Newtonian particle mechanics. Amusingly, this analogy is true even
in the details, for Schwarzschild solutions are mapped to
straight-line trajectories of free motion in flat superspace!

The arena for quantum gravity is not space-time but
superspace\cite{DeWitt,Wheeler}. Here we discover that the
corresponding minisuperspace is free of singularity and has a
surprisingly simple structure. We complete the Wheeler-DeWitt
canonical quantization program for the case at hand by using a
rather conservative approach with continuum variables, obtain the
complete solution to the quantum evolution with respect to
intrinsic superspace time, and discuss unitarity of the evolution
and the Hermiticity of the Hamiltonian in a clear and precise
manner. The contention that in all likelihood the mathematics of
our formalism is too special and restrictive to be applicable or
generalizable to the full theory is a point of view we can
empathize with; but the other side of the coin that black holes
can be so elegantly described (as free particles in Rindler
superpsace) suggests that the full theory may in fact contain
simplifications which have not yet been fully exploited, and that
our results and techniques may serve as launching points in
searching out the related exact states in the full theory and in
understanding their physical behavior.

The semiclassical limit from the quantum theory here (in a sense a
quantum derivation of Birkhoff's theorem) is in complete agreement
with Birkhoff's classical result of Schwarzschild solutions for
Einstein's field equations. All black holes (plane wave solutions
in our quantum context) are semiclassical regardless of the mass,
however small. Although this may be non-intuitive, the derived
classical limit is in fact reasonable, for all Schwarzschild
metrics regardless of mass are solutions of Einstein's equations
corresponding to stationary points of the gravitational action.
For these 4-mainfolds, black holes correspond to
(intrinsic)time-like straight line trajectories in superspace; and
for these trajectories intrinsic time $\tau$ and $R$ radial
coordinate time of interior Schwarzschild solutions are
monotonically related.

Classical singularities are precisely where the classical theory
breaks down. This means we cannot interpret the physics there by
drawing the correspondence between, even semiclassical, quantum
theory and the classical variables. The relevant question is
whether the quantum theory and the behavior of the quantum states
remain well defined. For the case at hand, quantum evolution takes
place in a non-singular Rindler wedge, and the semiclassical plane
wave states can be matched to Scwharzschild geometries, except at
$R=0$. Our formalism also shows that the coordinates $X_\pm$ are
well-defined even when $a,b,c$ diverge, and apparent classical
black hole ``singularities" are matched to finite values on the
$X_+$ axis. Moreover the classical EOM (9) also remains valid and
can, except for $R=0$, be interpreted in terms of Schwarzschild
geometry. What we lose is just the interpretation, only at $R=0$,
of the wavefunction as describing Schwarzchild black holes. This
is neither a heavy nor unreasonable price, for the result of
infinite curvature at $R=0$ also does not make sense even
classically.

While a ``physical observer" in a semiclassical Schwarzschild
background will have to face the issue of encountering the
``classical singularity" in a finite proper time, this situation
lies outside the scope of our work as we are only describing the
quantum evolution of pure spherically symmetric gravity without
matter, and for which the evolution is such that, for the plane
wave states, 3-geometries are stacked up in intrinsic time to
produce interior Schwarzschild 4-geometry. The demonstration of a
completely quantum description (albeit for spherically symmetric
sector) in the context of exact canonical quantum gravity leading
to the, from the physical point of view, interesting classical
limit of black hole physics while remaining at the same time a
unitary description with quantum evolution in intrinsic superspace
time even for situations with interior black hole geometry and
classical singularity is noteworthy. Although not pursued here, it
is possible to study the extension of the Rindler superspace to
the whole of Minkowski space-time and investigate ``singularity
traversal" and the continuation of the trajectories and their
correspondence to, in general, complex 4-manifolds.

Solutions of the Dirac equation associated with the Wheeler-DeWitt
constraint were also investigated. In flat Rindler space-time, a
solution of the former is also a solution of the latter. This is
not necessarily true if superspace is not flat in the full
theory\cite{DeWitt}. But the existence of fermionic solutions, and
``factorizations" of the Wheeler-DeWitt constraint, or its
generalizations with and without supersymmetry, should not be
ignored. The classical super-Hamiltonian constraint may be playing
the analogous role of a ``dispersion relation" which can be
realized by more than one type of ``particle" in the quantum
context. In the spherically symmetric sector our Klein-Gordon and
Dirac equations are really the resultant quantum constraints of a
massive free particle dispersion relation which is equivalent to
the classical super-Hamiltonian constraint of GR. For simplicity,
we have adopted a finite-dimensional representation of the Lorentz
group of superspace when the Dirac equation and its solutions were
discussed explicitly. Since superspace comes with hyperbolic
signature, the Lorentz group is non-compact and thus non-unitary
for finite-dimensional representations. This has the drawback of
the lack of unitary equivalence between states related by local
Lorentz transformations in superspace. In quantum field theory,
the issue is resolved by assuming that these wavefunctions are
operators acting upon physical states which belong to unitary
infinite-dimensional representations. In the present context such
a route would result in ``third quantization". Barring this, the
solutions here should already be considered as physical states of
the theory; however, one is not forbidden to consider these
solutions of the Dirac equation with infinite-dimensional unitary
representations.

There is a general belief that black holes are rudimentary objects
in GR\footnote{``Black holes as elementary particles" has been
conjectured many years ago, by G. 't Hooft for example (see, for
instance, {\it Holographic Views of the World} by G. Arcioni,
Sebastian de Haro and L. Susskind, in Contribution to Gerard 't
Hooft's 60th birthday celebration ``Under the Spell of Physics''
arXiv:physics/0611143).}. Intriguingly, this is bolstered by the
explicit correspondence that is established here between the
description of Schwarzschild black holes and that of elementary
free particle mechanics in superspace. Thus the conjecture that
black holes are elementary objects in classical and quantum
gravity, and also in particle physics, may in fact be more
accurate and noteworthy than expected.

\begin{acknowledgments}
This work has been supported in part by Grant No.
NSC95-2112-M-006-011-MY3 and the National Center for Theoretical
Sciences of Taiwan.

\end{acknowledgments}

\end{document}